\documentclass[prc,twocolumn,amsmath,amsfonts,aps,nofootinbib,preprintnumbers,letterpaper]{revtex4}
\usepackage{graphicx}
\usepackage{amssymb}
\usepackage{epsfig}

\newcommand\beq{\begin{eqnarray}}
\newcommand\eeq{\end{eqnarray}}

\newcommand\Fig[1]{Fig.~\ref{fig:#1}}
\newcommand\bal{\begin{align}}
\newcommand\eal{\end{align} }
\newcommand\eq[1]{Eq.~\ref{eq:#1}}

\newcommand\refcite[1]{Ref.~\cite{#1}}
\newcommand\refscite[1]{Refs.~\cite{#1}}

\begin{document}
\title{Lattice QCD Calculation of Nuclear Parity Violation}
\author{Joseph Wasem}\email{wasem2@llnl.gov} \affiliation{Lawrence Livermore National Laboratory, L-414, 7000 East Ave., Livermore, CA 94550, USA}

\preprint{LLNL-JRNL-491671}
\preprint{NT-UW-11-13}

\begin{abstract}
We present the first lattice QCD calculation of the leading-order momentum-independent parity violating coupling between pions and nucleons, $h_{\pi{NN}}^1$. The calculation performs measurements on dynamical anisotropic clover gauge configurations, with a spatial extent of $L\sim2.5$ fm, a spatial lattice spacing of $a_s\sim0.123$ fm, and a pion mass of $m_{\pi}\sim389$ MeV. While this first calculation does not include non-perturbative renormalization of the bare parity-violating operators, a chiral extrapolation to the physical pion mass, or contributions from disconnected (quark-loop) diagrams, these are expected to result in systematic errors within the quoted statistical error. We find a contribution from the `connected' diagrams of $h_{\pi{NN}}^{1,con}=(1.099\pm0.505^{+0.058}_{-0.064})\times10^{-7}$, consistent with current experimental bounds and previous model-dependent theoretical predictions.
\end{abstract}

\maketitle

Quantum chromodynamics (QCD) is the fundamental field theory that describes the dynamics and interactions of quarks and gluons, and the combination of QCD and electroweak interactions underlies all of nuclear physics. However, a quantitative understanding of nuclear observables directly from QCD has proved elusive due to the nonperturbative nature of the theory at low energies. Lattice QCD remains the sole avenue for theoretical explorations of observables in the nonperturbative regime with quantifiable errors. This is particularly meaningful for processes which are poorly understood experimentally, such as the neutral current parity violating (PV) weak interaction between quarks, which is the least understood portion of the standard model. In this work we report on the first calculation directly from QCD of the leading-order momentum-independent parity violating coupling between pions and nucleons, $h_{\pi{NN}}^{1}$, using $n_f=2+1$ lattice QCD calculations on configurations with a pion mass of $m_\pi\sim389$ MeV.

Parity violating interactions have been known since the late 1950s\cite{Lee:1956qn,Wu:1957my,Garwin:1957hc}, and their discovery radically changed perceptions of the role of fundamental symmetries in particle physics. While these interactions can be studied in flavor-changing decays, the effects of the PV neutral-current in such decays are tiny as the tree-level coupling between quarks and the $Z$ boson are flavor diagonal and radiative corrections are suppressed by the GIM mechanism\cite{Adelberger:1985ik,Glashow:1970gm}. This leaves PV flavor conserving interactions as the only laboratories for studying the weak neutral current, with the nucleon-nucleon (NN) PV interaction as the only accessible case. Isolation of the hadronic weak neutral current occurs in the $\Delta{I}=1$ NN channel, and this component is thought to be dominated by long-range pion exchange\cite{Adelberger:1985ik,Kaplan:1992vj}.

\begin{figure}[!h]
\centering
  \includegraphics[scale=.55]{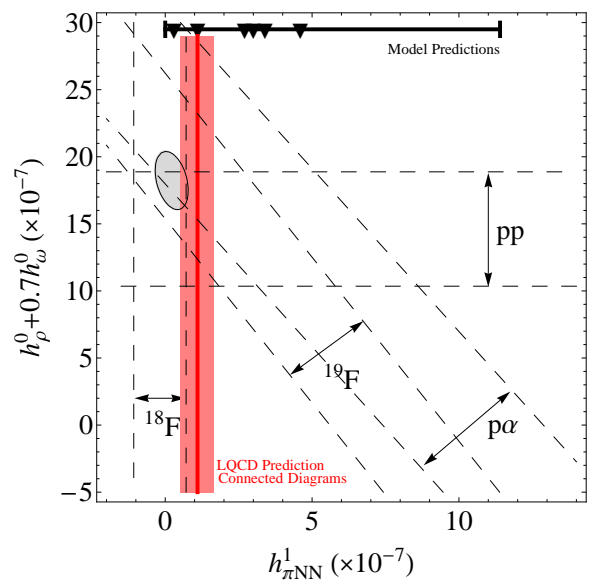}
  \caption[Experimental and model estimates of $h_{\pi{NN}}^{1}$ to date]{(Color online) Model estimates\cite{Kaiser:1988bt,Kaiser:1989fd,Dubovik:1986pj,Feldman:1991tj,Henley:1995ad,Henley:1998xh,Lobov:2002xb,Desplanques:1979hn} (solid line and triangles at top) and experimental results (dashed lines with labels and 1$\sigma$ error ellipse in grey, from \refscite{Adelberger:1985ik,Haeberli:1995uz,Haxton:2008ci} and references therein) for $h_{\pi{NN}}^{1}$ versus the dominant isoscalar PV coupling combination, along with the results of this work (solid vertical line and error band).}\label{fig:PVexpdata}
\end{figure}

At hadronic scales the weak interaction can be considered as a pointlike four-quark interaction which gives rise to a pion that mediates long-range interactions. Experiments to uncover this effect are technically demanding however, as the ratio of the weak to strong contributions to the NN interaction is approximately $10^{-7}$. In the decades since the discovery of parity violation, a heroic series of experiments (see \refscite{Adelberger:1985ik,Haeberli:1995uz,Haxton:2008ci} and references therein) have sought to uncover the value of $h_{\pi{NN}}^{1}$, defined in modern effective field theory language by\cite{Kaplan:1992vj}
\beq\label{eq:LPVhadronic}
    \mathcal{L}_{PV}^{\pi{NN}}&=&h_{\pi{NN}}^{1}\left(\bar{p}\pi^{+}n-\bar{n}\pi^{-}p\right)
\eeq
with a proton field $p$, neutron field $n$, and pion fields $\pi^{+}$/$\pi^{-}$. The most precise of these experiments are plotted with dashed lines in \Fig{PVexpdata}, with the combined $1\sigma$ error ellipse shown in grey. The coupling $h_{\pi{NN}}^{1}$ dominates the long range parity violating NN potential as it is not suppressed by powers of momentum. Although lacking precision, experimental results thus far suggest that while the isoscalar PV interaction is of natural size, the isovector interaction $h_{\pi{NN}}^{1}$ is suppressed. Early results from the most recent experimental collaboration to examine nuclear parity violation, the NPDGamma collaboration\cite{Gericke:2011zz}, have thus far not provided any significant constraint on $h_{\pi{NN}}^{1}$. However, the experiment is currently being reinstalled at the Spallation Neutron Source at Oak Ridge National Laboratory and should soon be able to reach its design precision.

Because QCD is nonperturbative, how the PV four-quark interactions build up into the composite interactions of the hadrons is not analytically known. Several model-dependent attempts have been made to calculate $h_{\pi{NN}}^{1}$ in such a way that the nonperturbative effects are included. The earliest of these used the quark model and symmetry considerations to make the first theoretical predictions of $h_{\pi{NN}}^{1}$\cite{Desplanques:1979hn} (the DDH result). Despite tremendous effort, the remaining systematic uncertainties from the nonperturbative sector of QCD prevented \refcite{Desplanques:1979hn} from specifying a result, and instead the outcome of the calculation was presented as a `best guess' with an accompanying range of values. Subsequent calculations using the quark model\cite{Dubovik:1986pj,Feldman:1991tj}, chiral solitons\cite{Kaiser:1988bt,Kaiser:1989fd}, and QCD sum rules\cite{Henley:1995ad,Henley:1998xh,Lobov:2002xb} have obtained greatly varying values of $h_{\pi{NN}}^{1}$, but all have remained within the original DDH range. The DDH range and the results of each model calculation are shown at the top of \Fig{PVexpdata}.

The lattice QCD calculation presented here uses anisotropic clover gauge configurations with 2 light quark flavors (the $u$ and $d$ quarks in the isospin limit, $m_u=m_d$) and one heavier quark flavor (the $s$ quark), at a pion mass of $389$ MeV, spatial lattice spacing of $0.123$ fm, and a temporal lattice spacing of $0.035$ fm\cite{Lin:2008pr,Edwards:2008ja}. The lattices have total dimensions of $(2.5$ fm$)^3\times9$ fm. Three-point correlation functions of the form
\begin{equation}\label{eq:3ptfn}
C_{A\to{B}}^{ij}(t,t')=\langle0|\mathcal{O}_{B,j}(t)\mathcal{O}_{PV}^{\Delta{I}=1}(t')\mathcal{O}^{\dagger}_{A,i}(0)|0\rangle
\end{equation}
are constructed, with $t$ the sink timeslice and $t'$ the operator insertion timeslice. In \eq{3ptfn}, the interpolating operator $\mathcal{O}^{\dagger}_{A,i}$ ($\mathcal{O}_{B,j}$) is used to create the initial state A (destroy final state B) with the quantum numbers of either the proton or the neutron-pion. The proton operator is $\epsilon^{abc}u_{a}(d^{T}_{b}C\gamma_5u_{c})$, with color indices $a$, $b$, $c$. Similarly the operator $\epsilon^{abc}\gamma_5u_{a}(d^{T}_{b}C\gamma_5u_{c})$ creates a neutron-pion state ($n\pi$) in an S-wave\cite{Beane:2009kya,Mahbub:2010me,Engel:2010my,Edwards:2011jj}.

Using a three-quark interpolating operator to create the $n\pi$ state greatly simplifies the contractions, and removes the need to calculate expensive quark-loop contributions at the sink which would arise from separate $n$ and $\pi$ operators. For large Euclidean times $t'$ and $t-t'$ the higher energy states induced by these interpolating operators will decay away (as determined from analysis of the two point functions of the form $C_{A,B}(t)=\langle0|\mathcal{O}_{A,B}(t)\mathcal{O}^{\dagger}_{A,B}(0)|0\rangle$), leaving only the proton or $n\pi$ S-wave state desired. Sandwiched between these operators in \eq{3ptfn} is the four-quark operator for the $\Delta{I}=1$ PV interaction. A ratio of the three-point functions to a combination of the two-point functions will plateau to the constant value of the desired parity violating matrix element.

The four-quark $\Delta{I}=1$ PV operator can be constructed directly from the standard electroweak interaction Lagrangian\cite{Glashow:1970gm} at the scale of the weak gauge bosons by integrating out the $Z$ boson (the contributions from the exchange of the $W^{\pm}$ bosons are neglected as they are suppressed by ${\rm sin}^2(\theta_C)\approx0.05$, where $\theta_C$ is the Cabibbo angle). One can then use continuum one-loop QCD perturbation theory to run the operator coefficients to the scale of the hadronic interactions ($\Lambda_\chi=1$ GeV) integrating out the heavier $b$- and $c$-quarks along the way\cite{Dai:1991bx,Kaplan:1992vj}. During the course of this running mixing between operators with the same quantum numbers will occur, leaving a total of 8 operators at the hadronic scale. There is no mixing with lower-dimension operators as the $\Delta{I}=1$ PV operator also \emph{conserves} CP, precluding quark bilinear operators from contributing with divergent inverse powers of the lattice spacing. The full four-quark $\Delta{I}=1$ PV operator at the hadronic scale can then be expressed as
\begin{equation}\label{eq:wkop}
\mathcal{O}_{PV}^{\Delta{I}=1}=-\frac{G_{F}{\rm sin}^2(\theta_W)}{3\sqrt{2}}\sum_{i=1}^{4}\int{d^3x}\left(C_{i}\theta_{i}^{q}+S_{i}\theta_{i}^{s}\right)
\end{equation}
where $G_F=1.16637\times10^{-5} \ {\rm GeV}^{-2}$ is the Fermi coupling and ${\rm sin}^2(\theta_W)=0.231$ is the weak mixing angle\cite{Nakamura:2010zzi}. The four-quark operators that contain only light ($u$ and $d$) quarks are $\theta_{i}^{q}$, while the $\theta_{i}^{s}$ contain $s$-quarks along with light quarks. The coefficients $C_{i}$ and $S_{i}$ of these operators and the specific operator forms used for $\theta_i$ in this work can be found in \refcite{Beane:2002ca}.

\begin{figure}[!h]
\centering
  \includegraphics[scale=.7]{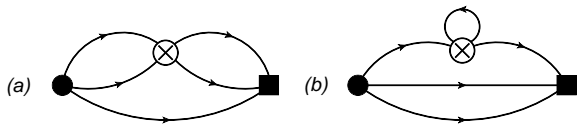}
  \caption[Contractions of the parity-violating operator]{The (a) connected and (b) quark-loop diagrams that contract the parity-violating operator with the interpolating operators for the source and sink. The filled circle and square represent the three-quark interpolating operators used at the source and the sink respectively, with one positive party and the other negative parity.}\label{fig:PVcontract}
\end{figure}

Performing the quark contractions in the three-point correlation function of \eq{3ptfn} using the above operators, one arrives at three possible diagrams for the quark propagators. The first type connects two of the quarks from both the source and sink operators to the weak operator, with the third quark going directly between the source and sink. This type is drawn in \Fig{PVcontract}(a) and is called the `connected' case. The second, `quark-loop' type of \Fig{PVcontract}(b) contains a quark loop at the weak operator insertion while connecting only one quark each from the source and sink to the weak operator. The final type contains a weak operator where all four quarks are contracted with each other, leading to an entirely `disconnected' contribution. However, in the isospin limit the contributions from this type of diagram will sum to zero, saving considerable computational expense. Because the interpolating operators consist entirely of light quarks, the operators $\theta_{i}^{q}$ will have contributions to both the connected and quark-loop diagrams, while the operators $\theta_{i}^{s}$ will contribute only to the quark-loop diagrams as the $s$-quarks will be required to be contained in the quark loop itself.

Typically 3-point correlation functions are computed on the lattice using an efficient technique known as sequential inversion, whereby the quark propagators calculated from the source to the sink are contracted into a new `source' which is inverted to obtain the propagator backwards to the operator insertion. However, this technique fails for this calculation both in the case of the connected diagrams (due to the need for two propagators between the operator and the sink) and in the case of the quark loop diagrams (as the quark-loop would remain to be calculated). Instead, this calculation performs two separate quark propagator inversions, one at the source and one at the weak operator insertion. This method unfortunately restricts the measurements to a single spatial site on the operator timeslice (all spatial sites are sampled over the course of the calculation), but allows for maximum flexibility and computational efficiency (as the propagators may be used for both the connected and the quark-loop diagrams, and for any of the weak operators). With this method, the timeslice on which the weak operator is placed ($t'$) must be large enough that the excited states of the source operator are exponentially small, and for this calculation $t'=24$.

As previously mentioned, to extract the desired matrix element a ratio of 3-point and 2-point functions must be formed such that in the limit of large $t'$ and $t-t'$ contamination from excited states dies off and the ground state overlap factors are canceled, allowing the ratio to plateau to the value of the matrix element. This ratio is given by
\begin{equation}\label{eq:ratio}
R_{A\to{B}}^{ij}=\frac{C_{A\to{B}}^{ij}(t,t')}{C_{B}^{jj}(t)}\left(\frac{C_{A}^{ii}(t-t')C_{B}^{jj}(t)C_{B}^{jj}(t')}{C_{B}^{jj}(t-t')C_{A}^{ii}(t)C_{A}^{ii}(t')}\right)^{\frac{1}{2}}
\end{equation}
where the smearings $i$ and $j$ used in the 2-point functions must match that used for the corresponding state in the 3-point function in order to have the correct cancelation of overlap factors. However, as discussed in \refcite{Beane:2002ca} the differing energy levels of the proton and the $n\pi$ states will cause an insertion of energy by the weak operator to occur, modifying \eq{LPVhadronic} to
\begin{equation}\label{eq:LPVhadtotal}
    \mathcal{L}_{PV}^{\pi{NN}}=h_{\pi{NN}}^{1}\left(\bar{p}\pi^{+}n-\bar{n}\pi^{-}p\right)+h_{E}D_{t}\left(\bar{p}\pi^{+}n-\bar{n}\pi^{-}p\right)
\end{equation}
with some unknown coefficient $h_{E}$, making the long-time behavior of \eq{ratio}
\begin{eqnarray}\label{eq:ratlim}
R_{p\to{n\pi}}^{ij}&\to&{h}_{\pi{NN}}^{1}+\left[E_{n\pi}-E_{p}\right]\cdot{h_{E}}\nonumber\\
R_{{n\pi}\to{p}}^{ij}&\to&{-}\left(h_{\pi{NN}}^{1}+\left[E_{p}-E_{n\pi}\right]\cdot{h_{E}}\right)
\end{eqnarray}
with $E_{p}$ and $E_{n\pi}$ the energy levels of the proton and $n\pi$ states, respectively. However, while the energy injection term is present in both the forward ($p\to{n\pi}$) and backward ($n\pi\to{p}$) interactions, it can be eliminated with an antisymmetric combination of \eq{ratlim}, leading to a plateau region given by
\begin{equation}\label{eq:rathpnn}
H^{ij}=\frac{1}{2}\left(R_{p\to{n\pi}}^{ij}-R_{n\pi\to{p}}^{ij}\right)\to{h}_{\pi{NN}}^{1}.
\end{equation}

\begin{figure}[!h]
\centering
  \includegraphics[scale=.55]{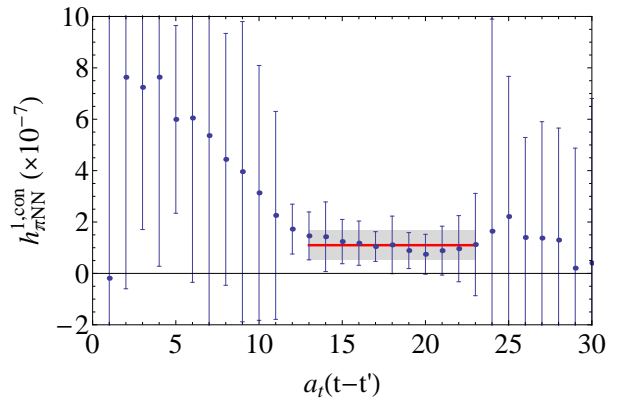}
  \caption[$h_{\pi{NN}}^{1}(\theta_{con})$ Plateau]{(Color online) Lattice results for the contribution of connected quark diagrams to $h_{\pi{NN}}^{1}$, as a function of Euclidean lattice time from the operator insertion. The solid line is the fully correlated fit value over the plateau region with the grey rectangle the statistical plus fit window systematic uncertainty.}\label{fig:PVplat}
\end{figure}

A total of 100,871 measurements of each of the smearing combinations of $H^{ij}$ are performed, where $i,j$ can be either point- or shell-smearing. These measurements are then blocked on each configuration and bootstrapped. One can enhance the plateau region for \eq{rathpnn} by taking appropriately normalized linear combinations of the different smearing combinations, using the matrix-prony\cite{Beane:2009gs} method  on the bootstrapped ensemble to determine the optimal linear combination. This is done for both the connected and the quark-loop contractions. In the case of the quark-loop diagrams the signal-to-noise ratio remains far too small to recover any reliable result, and we do not attempt to extract a signal. It is expected that improvements in both contraction algorithms and overall calculation runtime will be needed to overcome this difficulty and reliably extract the quark-loop contribution. For the connected contributions the analysis returns the data shown in \Fig{PVplat}, revealing not only a well defined plateau region, but a robust non-zero contribution to $h_{\pi{NN}}^{1}$.

In \Fig{PVplat}, a fully correlated $\chi^2$ minimizing fit to a constant is performed over the plateau region, with additional systematic error due to the choice of plateau region determined by shifting the ends of the region $\pm2$ timeslices. The quoted systematic error is one-half of the maximum minus the minimum of these shifted fits. The fit result and statistical plus systematic error are shown in \Fig{PVplat} with the solid line and grey band. The contribution of the connected diagrams to $h_{\pi{NN}}^{1}$ is then found to be
\begin{equation}\label{eq:result}
h_{\pi{NN}}^{1,con}=(1.099\pm0.505^{+0.058}_{-0.064})\times10^{-7}
\end{equation}
where the first and second uncertainties are statistical and systematic, respectively. The fit result is plotted in \Fig{PVexpdata} as the vertical line and error band, and it is consistent with both experimental bounds and previous model calculations.

Nonperturbative renormalization of the bare PV operators at the lattice scale and subsequent matching to a perturbative scheme is not performed for this first calculation, though results from other four-quark calculations (at similar pion mass and lattice spacing) indicate that this should affect the result by a value significantly below the quoted statistical error\cite{Blum:2011pu}. With the clover action, lattice spacing errors are expected to be $\mathcal{O}(a_s^2\Lambda_{QCD}^2)\sim2\%$, also well below statistical uncertainty. Because sequential propagators are not used, one largely eliminates excited state contamination by choosing an operator insertion time well into the 2-point correlation function ground state plateaus (though this necessarily increases the statistical uncertainty by pushing the sink operator further into the baryon noise). Finally, one expects from chiral perturbation theory the next-to-leading-order contribution to be a pion loop originating at the operator insertion, giving an expected finite volume error of $\mathcal{O}((m_{\pi}f_{\pi}^2L^3)^{-1})\sim7\%$ ($f_\pi=132$ MeV is the pion decay constant). While future calculations must also address these sources of systematic error, the uncertainties in this work remain dominated by statistical uncertainty.

In conclusion, we have performed the first calculation of the quantity $h_{\pi{NN}}^{1}$ directly from the underlying theory of QCD. Our calculation was performed on one ensemble of anisotropic clover configurations with a pion mass of $m_{\pi}\sim389$ MeV. Future calculations will need to be performed at pion masses closer to the physical point, with sufficient statistical resolution to extract the contribution of the quark-loop diagrams (expected to be on the order of $10^3$x more measurements), and include non-perturbative renormalization of the included operators. While significant technical challenges remain in the calculation of the full matrix element, this first of its kind result clearly shows that lattice QCD can make a significant contribution to the theoretical, model-independent, understanding of quantities that are difficult to access experimentally. Our initial result shows good agreement with current experimental bounds and paves the way toward a complete extraction of $h_{\pi{NN}}^{1}$ at a precision consistent with, or better than, the anticipated results of the upcoming NPDGamma experiment at Oak Ridge.

We thank M. J. Savage, T. Luu, A. Nicholson, B. Smigielski, A. Walker-Loud, B. Tiburzi, M. Buchoff, W. Haxton and P. Vranas for many useful discussions and B. Joo for help with QDP++ and CHROMA\cite{Edwards:2004sx}. Computations were performed on the uBGL and Edge clusters at LLNL. This work was performed under the auspices of the U.S. Department of Energy by LLNL under Contract No. DE-AC52-07NA27344 and the UNEDF SciDAC Grant No. DE-FC02-07ER41457. Early work at the University of Washington was performed under DOE Grant No. DE-FG02-97ER41014.

\bibliography{masterbib}

\begin{thebibliography}{29}
\expandafter\ifx\csname natexlab\endcsname\relax\def\natexlab#1{#1}\fi
\expandafter\ifx\csname bibnamefont\endcsname\relax
  \def\bibnamefont#1{#1}\fi
\expandafter\ifx\csname bibfnamefont\endcsname\relax
  \def\bibfnamefont#1{#1}\fi
\expandafter\ifx\csname citenamefont\endcsname\relax
  \def\citenamefont#1{#1}\fi
\expandafter\ifx\csname url\endcsname\relax
  \def\url#1{\texttt{#1}}\fi
\expandafter\ifx\csname urlprefix\endcsname\relax\def\urlprefix{URL }\fi
\providecommand{\bibinfo}[2]{#2}
\providecommand{\eprint}[2][]{\url{#2}}

\bibitem[{\citenamefont{Lee and Yang}(1956)}]{Lee:1956qn}
\bibinfo{author}{\bibfnamefont{T.~D.} \bibnamefont{Lee}} \bibnamefont{and}
  \bibinfo{author}{\bibfnamefont{C.-N.} \bibnamefont{Yang}},
  \bibinfo{journal}{Phys. Rev.} \textbf{\bibinfo{volume}{104}},
  \bibinfo{pages}{254} (\bibinfo{year}{1956}).

\bibitem[{\citenamefont{Wu et~al.}(1957)\citenamefont{Wu, Ambler, Hayward,
  Hoppes, and Hudson}}]{Wu:1957my}
\bibinfo{author}{\bibfnamefont{C.}~\bibnamefont{Wu}},
  \bibinfo{author}{\bibfnamefont{E.}~\bibnamefont{Ambler}},
  \bibinfo{author}{\bibfnamefont{R.}~\bibnamefont{Hayward}},
  \bibinfo{author}{\bibfnamefont{D.}~\bibnamefont{Hoppes}}, \bibnamefont{and}
  \bibinfo{author}{\bibfnamefont{R.}~\bibnamefont{Hudson}},
  \bibinfo{journal}{Phys.Rev.} \textbf{\bibinfo{volume}{105}},
  \bibinfo{pages}{1413} (\bibinfo{year}{1957}).

\bibitem[{\citenamefont{Garwin et~al.}(1957)\citenamefont{Garwin, Lederman, and
  Weinrich}}]{Garwin:1957hc}
\bibinfo{author}{\bibfnamefont{R.~L.} \bibnamefont{Garwin}},
  \bibinfo{author}{\bibfnamefont{L.~M.} \bibnamefont{Lederman}},
  \bibnamefont{and} \bibinfo{author}{\bibfnamefont{M.}~\bibnamefont{Weinrich}},
  \bibinfo{journal}{Phys. Rev.} \textbf{\bibinfo{volume}{105}},
  \bibinfo{pages}{1415} (\bibinfo{year}{1957}).

\bibitem[{\citenamefont{Adelberger and Haxton}(1985)}]{Adelberger:1985ik}
\bibinfo{author}{\bibfnamefont{E.~G.} \bibnamefont{Adelberger}}
  \bibnamefont{and} \bibinfo{author}{\bibfnamefont{W.~C.}
  \bibnamefont{Haxton}}, \bibinfo{journal}{Ann. Rev. Nucl. Part. Sci.}
  \textbf{\bibinfo{volume}{35}}, \bibinfo{pages}{501} (\bibinfo{year}{1985}).

\bibitem[{\citenamefont{Glashow et~al.}(1970)\citenamefont{Glashow, Iliopoulos,
  and Maiani}}]{Glashow:1970gm}
\bibinfo{author}{\bibfnamefont{S.~L.} \bibnamefont{Glashow}},
  \bibinfo{author}{\bibfnamefont{J.}~\bibnamefont{Iliopoulos}},
  \bibnamefont{and} \bibinfo{author}{\bibfnamefont{L.}~\bibnamefont{Maiani}},
  \bibinfo{journal}{Phys. Rev.} \textbf{\bibinfo{volume}{D2}},
  \bibinfo{pages}{1285} (\bibinfo{year}{1970}).

\bibitem[{\citenamefont{Kaplan and Savage}(1993)}]{Kaplan:1992vj}
\bibinfo{author}{\bibfnamefont{D.~B.} \bibnamefont{Kaplan}} \bibnamefont{and}
  \bibinfo{author}{\bibfnamefont{M.~J.} \bibnamefont{Savage}},
  \bibinfo{journal}{Nucl. Phys.} \textbf{\bibinfo{volume}{A556}},
  \bibinfo{pages}{653} (\bibinfo{year}{1993}).

\bibitem[{\citenamefont{Kaiser and Meissner}(1988)}]{Kaiser:1988bt}
\bibinfo{author}{\bibfnamefont{N.}~\bibnamefont{Kaiser}} \bibnamefont{and}
  \bibinfo{author}{\bibfnamefont{U.~G.} \bibnamefont{Meissner}},
  \bibinfo{journal}{Nucl. Phys.} \textbf{\bibinfo{volume}{A489}},
  \bibinfo{pages}{671} (\bibinfo{year}{1988}).

\bibitem[{\citenamefont{Kaiser and Meissner}(1989)}]{Kaiser:1989fd}
\bibinfo{author}{\bibfnamefont{N.}~\bibnamefont{Kaiser}} \bibnamefont{and}
  \bibinfo{author}{\bibfnamefont{U.~G.} \bibnamefont{Meissner}},
  \bibinfo{journal}{Nucl. Phys.} \textbf{\bibinfo{volume}{A499}},
  \bibinfo{pages}{699} (\bibinfo{year}{1989}).

\bibitem[{\citenamefont{Dubovik and Zenkin}(1986)}]{Dubovik:1986pj}
\bibinfo{author}{\bibfnamefont{V.~M.} \bibnamefont{Dubovik}} \bibnamefont{and}
  \bibinfo{author}{\bibfnamefont{S.~V.} \bibnamefont{Zenkin}},
  \bibinfo{journal}{Annals Phys.} \textbf{\bibinfo{volume}{172}},
  \bibinfo{pages}{100} (\bibinfo{year}{1986}).

\bibitem[{\citenamefont{Feldman et~al.}(1991)\citenamefont{Feldman, Crawford,
  Dubach, and Holstein}}]{Feldman:1991tj}
\bibinfo{author}{\bibfnamefont{G.}~\bibnamefont{Feldman}},
  \bibinfo{author}{\bibfnamefont{G.~A.} \bibnamefont{Crawford}},
  \bibinfo{author}{\bibfnamefont{J.}~\bibnamefont{Dubach}}, \bibnamefont{and}
  \bibinfo{author}{\bibfnamefont{B.~R.} \bibnamefont{Holstein}},
  \bibinfo{journal}{Phys.Rev.} \textbf{\bibinfo{volume}{C43}},
  \bibinfo{pages}{863} (\bibinfo{year}{1991}).

\bibitem[{\citenamefont{Henley et~al.}(1996)\citenamefont{Henley, Hwang, and
  Kisslinger}}]{Henley:1995ad}
\bibinfo{author}{\bibfnamefont{E.~M.} \bibnamefont{Henley}},
  \bibinfo{author}{\bibfnamefont{W.~Y.~P.} \bibnamefont{Hwang}},
  \bibnamefont{and} \bibinfo{author}{\bibfnamefont{L.~S.}
  \bibnamefont{Kisslinger}}, \bibinfo{journal}{Phys. Lett.}
  \textbf{\bibinfo{volume}{B367}}, \bibinfo{pages}{21} (\bibinfo{year}{1996}),
  \eprint{nucl-th/9511002}.

\bibitem[{\citenamefont{Henley et~al.}(1998)\citenamefont{Henley, Hwang, and
  Kisslinger}}]{Henley:1998xh}
\bibinfo{author}{\bibfnamefont{E.~M.} \bibnamefont{Henley}},
  \bibinfo{author}{\bibfnamefont{W.~Y.~P.} \bibnamefont{Hwang}},
  \bibnamefont{and} \bibinfo{author}{\bibfnamefont{L.~S.}
  \bibnamefont{Kisslinger}}, \bibinfo{journal}{Phys. Lett.}
  \textbf{\bibinfo{volume}{B440}}, \bibinfo{pages}{449} (\bibinfo{year}{1998}),
  \eprint{nucl-th/9809064}.

\bibitem[{\citenamefont{Lobov}(2002)}]{Lobov:2002xb}
\bibinfo{author}{\bibfnamefont{G.~A.} \bibnamefont{Lobov}},
  \bibinfo{journal}{Phys. Atom. Nucl.} \textbf{\bibinfo{volume}{65}},
  \bibinfo{pages}{534} (\bibinfo{year}{2002}).

\bibitem[{\citenamefont{Desplanques et~al.}(1980)\citenamefont{Desplanques,
  Donoghue, and Holstein}}]{Desplanques:1979hn}
\bibinfo{author}{\bibfnamefont{B.}~\bibnamefont{Desplanques}},
  \bibinfo{author}{\bibfnamefont{J.~F.} \bibnamefont{Donoghue}},
  \bibnamefont{and} \bibinfo{author}{\bibfnamefont{B.~R.}
  \bibnamefont{Holstein}}, \bibinfo{journal}{Ann. Phys.}
  \textbf{\bibinfo{volume}{124}}, \bibinfo{pages}{449} (\bibinfo{year}{1980}).

\bibitem[{\citenamefont{Haeberli and Holstein}(1995)}]{Haeberli:1995uz}
\bibinfo{author}{\bibfnamefont{W.}~\bibnamefont{Haeberli}} \bibnamefont{and}
  \bibinfo{author}{\bibfnamefont{B.~R.} \bibnamefont{Holstein}},
  \bibinfo{journal}{Symmetries and fundamental interactions in nuclei} pp.
  \bibinfo{pages}{17--66} (\bibinfo{year}{1995}).

\bibitem[{\citenamefont{Haxton}(2008)}]{Haxton:2008ci}
\bibinfo{author}{\bibfnamefont{W.~C.} \bibnamefont{Haxton}}
  (\bibinfo{year}{2008}), \eprint{arXiv:0802.2984 [nucl-th]}.

\bibitem[{\citenamefont{Gericke et~al.}(2011)}]{Gericke:2011zz}
\bibinfo{author}{\bibfnamefont{M.}~\bibnamefont{Gericke}} \bibnamefont{et~al.},
  \bibinfo{journal}{Phys.Rev.} \textbf{\bibinfo{volume}{C83}},
  \bibinfo{pages}{015505} (\bibinfo{year}{2011}).

\bibitem[{\citenamefont{Lin et~al.}(2009)\citenamefont{Lin, Cohen, Dudek,
  Edwards, Joo, Richards, Bulava, Foley, Morningstar, Engelson
  et~al.}}]{Lin:2008pr}
\bibinfo{author}{\bibfnamefont{H.-W.} \bibnamefont{Lin}},
  \bibinfo{author}{\bibfnamefont{S.~D.} \bibnamefont{Cohen}},
  \bibinfo{author}{\bibfnamefont{J.}~\bibnamefont{Dudek}},
  \bibinfo{author}{\bibfnamefont{R.~G.} \bibnamefont{Edwards}},
  \bibinfo{author}{\bibfnamefont{B.}~\bibnamefont{Joo}},
  \bibinfo{author}{\bibfnamefont{D.~G.} \bibnamefont{Richards}},
  \bibinfo{author}{\bibfnamefont{J.}~\bibnamefont{Bulava}},
  \bibinfo{author}{\bibfnamefont{J.}~\bibnamefont{Foley}},
  \bibinfo{author}{\bibfnamefont{C.}~\bibnamefont{Morningstar}},
  \bibinfo{author}{\bibfnamefont{E.}~\bibnamefont{Engelson}},
  \bibnamefont{et~al.} (\bibinfo{collaboration}{Hadron Spectrum
  Collaboration}), \bibinfo{journal}{Phys.Rev.} \textbf{\bibinfo{volume}{D79}},
  \bibinfo{pages}{034502} (\bibinfo{year}{2009}), \eprint{0810.3588}.

\bibitem[{\citenamefont{Edwards et~al.}(2008)\citenamefont{Edwards, Joo, and
  Lin}}]{Edwards:2008ja}
\bibinfo{author}{\bibfnamefont{R.~G.} \bibnamefont{Edwards}},
  \bibinfo{author}{\bibfnamefont{B.}~\bibnamefont{Joo}}, \bibnamefont{and}
  \bibinfo{author}{\bibfnamefont{H.-W.} \bibnamefont{Lin}},
  \bibinfo{journal}{Phys.Rev.} \textbf{\bibinfo{volume}{D78}},
  \bibinfo{pages}{054501} (\bibinfo{year}{2008}), \eprint{0803.3960}.

\bibitem[{\citenamefont{Beane et~al.}(2009{\natexlab{a}})\citenamefont{Beane,
  Detmold, Luu, Orginos, Parreno, Savage, Torok, and
  Walker-Loud}}]{Beane:2009kya}
\bibinfo{author}{\bibfnamefont{S.~R.} \bibnamefont{Beane}},
  \bibinfo{author}{\bibfnamefont{W.}~\bibnamefont{Detmold}},
  \bibinfo{author}{\bibfnamefont{T.~C.} \bibnamefont{Luu}},
  \bibinfo{author}{\bibfnamefont{K.}~\bibnamefont{Orginos}},
  \bibinfo{author}{\bibfnamefont{A.}~\bibnamefont{Parreno}},
  \bibinfo{author}{\bibfnamefont{M.~J.} \bibnamefont{Savage}},
  \bibinfo{author}{\bibfnamefont{A.}~\bibnamefont{Torok}}, \bibnamefont{and}
  \bibinfo{author}{\bibfnamefont{A.}~\bibnamefont{Walker-Loud}},
  \bibinfo{journal}{Phys. Rev.} \textbf{\bibinfo{volume}{D79}},
  \bibinfo{pages}{114502} (\bibinfo{year}{2009}{\natexlab{a}}),
  \eprint{0903.2990}.

\bibitem[{\citenamefont{Mahbub et~al.}(2010)\citenamefont{Mahbub, Kamleh,
  Leinweber, O~Cais, and Williams}}]{Mahbub:2010me}
\bibinfo{author}{\bibfnamefont{M.}~\bibnamefont{Mahbub}},
  \bibinfo{author}{\bibfnamefont{W.}~\bibnamefont{Kamleh}},
  \bibinfo{author}{\bibfnamefont{D.~B.} \bibnamefont{Leinweber}},
  \bibinfo{author}{\bibfnamefont{A.}~\bibnamefont{O~Cais}}, \bibnamefont{and}
  \bibinfo{author}{\bibfnamefont{A.~G.} \bibnamefont{Williams}},
  \bibinfo{journal}{Phys.Lett.} \textbf{\bibinfo{volume}{B693}},
  \bibinfo{pages}{351} (\bibinfo{year}{2010}), \eprint{1007.4871}.

\bibitem[{\citenamefont{Engel et~al.}(2010)\citenamefont{Engel, Lang, Limmer,
  Mohler, and Schafer}}]{Engel:2010my}
\bibinfo{author}{\bibfnamefont{G.~P.} \bibnamefont{Engel}},
  \bibinfo{author}{\bibfnamefont{C.}~\bibnamefont{Lang}},
  \bibinfo{author}{\bibfnamefont{M.}~\bibnamefont{Limmer}},
  \bibinfo{author}{\bibfnamefont{D.}~\bibnamefont{Mohler}}, \bibnamefont{and}
  \bibinfo{author}{\bibfnamefont{A.}~\bibnamefont{Schafer}}
  (\bibinfo{collaboration}{BGR [Bern-Graz-Regensburg] Collaboration}),
  \bibinfo{journal}{Phys.Rev.} \textbf{\bibinfo{volume}{D82}},
  \bibinfo{pages}{034505} (\bibinfo{year}{2010}), \eprint{1005.1748}.

\bibitem[{\citenamefont{Edwards et~al.}(2011)\citenamefont{Edwards, Dudek,
  Richards, and Wallace}}]{Edwards:2011jj}
\bibinfo{author}{\bibfnamefont{R.~G.} \bibnamefont{Edwards}},
  \bibinfo{author}{\bibfnamefont{J.~J.} \bibnamefont{Dudek}},
  \bibinfo{author}{\bibfnamefont{D.~G.} \bibnamefont{Richards}},
  \bibnamefont{and} \bibinfo{author}{\bibfnamefont{S.~J.}
  \bibnamefont{Wallace}} (\bibinfo{year}{2011}), \eprint{1104.5152}.

\bibitem[{\citenamefont{Dai et~al.}(1991)\citenamefont{Dai, Savage, Liu, and
  Springer}}]{Dai:1991bx}
\bibinfo{author}{\bibfnamefont{J.}~\bibnamefont{Dai}},
  \bibinfo{author}{\bibfnamefont{M.~J.} \bibnamefont{Savage}},
  \bibinfo{author}{\bibfnamefont{J.}~\bibnamefont{Liu}}, \bibnamefont{and}
  \bibinfo{author}{\bibfnamefont{R.~P.} \bibnamefont{Springer}},
  \bibinfo{journal}{Phys. Lett.} \textbf{\bibinfo{volume}{B271}},
  \bibinfo{pages}{403} (\bibinfo{year}{1991}).

\bibitem[{\citenamefont{Nakamura et~al.}(2010)}]{Nakamura:2010zzi}
\bibinfo{author}{\bibfnamefont{K.}~\bibnamefont{Nakamura}} \bibnamefont{et~al.}
  (\bibinfo{collaboration}{Particle Data Group}), \bibinfo{journal}{J.Phys.G}
  \textbf{\bibinfo{volume}{G37}}, \bibinfo{pages}{075021}
  (\bibinfo{year}{2010}).

\bibitem[{\citenamefont{Beane and Savage}(2002)}]{Beane:2002ca}
\bibinfo{author}{\bibfnamefont{S.~R.} \bibnamefont{Beane}} \bibnamefont{and}
  \bibinfo{author}{\bibfnamefont{M.~J.} \bibnamefont{Savage}},
  \bibinfo{journal}{Nucl. Phys.} \textbf{\bibinfo{volume}{B636}},
  \bibinfo{pages}{291} (\bibinfo{year}{2002}), \eprint{hep-lat/0203028}.

\bibitem[{\citenamefont{Beane et~al.}(2009{\natexlab{b}})\citenamefont{Beane,
  Detmold, Luu, Orginos, Parreno, Savage, Torok, and
  Walker-Loud}}]{Beane:2009gs}
\bibinfo{author}{\bibfnamefont{S.~R.} \bibnamefont{Beane}},
  \bibinfo{author}{\bibfnamefont{W.}~\bibnamefont{Detmold}},
  \bibinfo{author}{\bibfnamefont{T.~C.} \bibnamefont{Luu}},
  \bibinfo{author}{\bibfnamefont{K.}~\bibnamefont{Orginos}},
  \bibinfo{author}{\bibfnamefont{A.}~\bibnamefont{Parreno}},
  \bibinfo{author}{\bibfnamefont{M.~J.} \bibnamefont{Savage}},
  \bibinfo{author}{\bibfnamefont{A.}~\bibnamefont{Torok}}, \bibnamefont{and}
  \bibinfo{author}{\bibfnamefont{A.}~\bibnamefont{Walker-Loud}},
  \bibinfo{journal}{Phys. Rev.} \textbf{\bibinfo{volume}{D80}},
  \bibinfo{pages}{074501} (\bibinfo{year}{2009}{\natexlab{b}}),
  \eprint{0905.0466}.

\bibitem[{\citenamefont{Blum et~al.}(2011)\citenamefont{Blum, Boyle, Christ,
  Garron, Goode, Izubuchi, Lehner, Liu, Mawhinney, Sachrjda
  et~al.}}]{Blum:2011pu}
\bibinfo{author}{\bibfnamefont{T.}~\bibnamefont{Blum}},
  \bibinfo{author}{\bibfnamefont{P.}~\bibnamefont{Boyle}},
  \bibinfo{author}{\bibfnamefont{N.}~\bibnamefont{Christ}},
  \bibinfo{author}{\bibfnamefont{N.}~\bibnamefont{Garron}},
  \bibinfo{author}{\bibfnamefont{E.}~\bibnamefont{Goode}},
  \bibinfo{author}{\bibfnamefont{T.}~\bibnamefont{Izubuchi}},
  \bibinfo{author}{\bibfnamefont{C.}~\bibnamefont{Lehner}},
  \bibinfo{author}{\bibfnamefont{Q.}~\bibnamefont{Liu}},
  \bibinfo{author}{\bibfnamefont{R.}~\bibnamefont{Mawhinney}},
  \bibinfo{author}{\bibfnamefont{C.}~\bibnamefont{Sachrjda}},
  \bibnamefont{et~al.}, \bibinfo{journal}{Phys.Rev.}
  \textbf{\bibinfo{volume}{D84}}, \bibinfo{pages}{114503}
  (\bibinfo{year}{2011}), \eprint{1106.2714}.

\bibitem[{\citenamefont{Edwards and Joo}(2005)}]{Edwards:2004sx}
\bibinfo{author}{\bibfnamefont{R.~G.} \bibnamefont{Edwards}} \bibnamefont{and}
  \bibinfo{author}{\bibfnamefont{B.}~\bibnamefont{Joo}}
  (\bibinfo{collaboration}{SciDAC}), \bibinfo{journal}{Nucl. Phys. Proc.
  Suppl.} \textbf{\bibinfo{volume}{140}}, \bibinfo{pages}{832}
  (\bibinfo{year}{2005}), \eprint{hep-lat/0409003}.

\end{thebibliography}

\end{document}